# Broadband Asymmetric Transmission with Wide Spectral Tunability based on Substrate-Embedded Silicon Nanoring Arrays

Ruihan Ma[1], Yuqing Cheng[1,*] and Mengtao Sun[1,†]

[1] School of Mathematics and Physics, University of Science and Technology Beijing, Beijing 100083, People's Republic of China

**Abstract:** In this work, we theoretically propose a broadband asymmetric transmission (AT) device based on periodic Si nanoring arrays embedded in a $SiO_2$ substrate. Results indicate that the device achieves a remarkable broadband AT effect in the near-infrared region (1750-2400 nm), with forward transmissivity exceeding 0.8 (maximum of 0.98), backward transmissivity less than 0.15 (minimum of 0.015) and an isolation ratio (*IR*) reaching a maximum of 17.8 dB at 2280 nm. Furthermore, the transmissivity spectrum exhibits excellent scalability and tunability through uniform scaling of the structure, allowing the operational band to be tailored across a wide spectral range, from 890 to 3300 nm. This Si-based nanostructure offers a robust and flexible platform for applications in optical isolation, multi-channel sensing, and integrated photonic circuits.

## 1. Introduction

Asymmetric transmission (AT) refers to the phenomenon in which light exhibits significantly different transmissivity under forward and backward incidence. This optical "diode-like" behavior holds great potential for applications in optical communications[1-3], integrated photonics[4, 5], optical isolation[6-8], and multi-channel sensing[9, 10]. Traditionally, optical AT has been realized based on magneto-optical effects, however, such approaches suffer from poor compatibility with Complementary metal-oxide-semiconductor (CMOS) fabrication processes and the requirement of external magnetic fields. In contrast, all-dielectric nanostructures that break spatial inversion symmetry provide an alternative route to achieve asymmetric optical transmission in reciprocal systems[11-15], and have attracted considerable attention due to their low-loss characteristics and excellent compatibility with on-chip integration. To date, various structures have been proposed to realize AT, including composite gratings[16, 17], photonic crystals[18, 19], chiral metamaterials[20-22], and all-dielectric metasurfaces[23-25]. Nevertheless, achieving simultaneously high efficiency and broadband performance in the visible and near-infrared regimes remains challenging. On one hand, metallic components introduce intrinsic Ohmic losses that significantly reduce transmissivity efficiency[26-29]; on the other hand, material dispersion limits the bandwidth over which high-performance operation can be maintained. Although silicon nanorod arrays have been explored in the visible regime, studies on silicon-based nanoring arrays operating in the near-infrared region with enhanced tunability are still relatively scarce[30, 31].

[*] Email: yuqingcheng@ustb.edu.cn
[†] Email: mengtaosun@ustb.edu.cn

In this work, we theoretically propose a broadband AT device in the near-infrared region based on a periodical silicon nanoring array embedded in a $SiO_2$ substrate. Furthermore, when uniformly scaling the size of the structure, the AT band can be shifted accordingly. This kind of structure provides a promising route toward high-performance and integrable passive optical isolation devices.

## 2. Method and structure

Fig. 1 illustrates the schematic of the periodic Si nanoring array embedded within a $SiO_2$ substrate. In this configuration, the array is defined by a rectangular lattice with periods $p_x$ and $p_y$ along the x- and y-directions, respectively, where the relationship $p_y = \sqrt{3}p_x$ is maintained, resulting in offset-periodic nanoring array. High-efficiency AT is achieved by optimizing the geometric parameters, the inner radius ($r_1$), outer radius ($r_2$), height ($h$), and period ($p_x$ and $p_y$) to maximize forward transmittance while simultaneously suppressing backward transmittance. To obtain the optimal AT performance at specific target wavelengths, these parameters are systematically determined through a sequential optimization process. The structural parameters of the optimized structure are summarized in Table 1. For all simulations and analysis, the upper surface of the substrate is defined as the z = 0 plane.

The optical response of the proposed device is numerically analyzed using the three-dimensional finite-difference time-domain (3D-FDTD) method. In the simulations, x-polarized plane waves are employed as the incident light, propagating along the −z (forward) and +z (backward) directions at normal incidence. Periodic boundary conditions (PBCs) are applied in the x and y directions to simulate the array environment, while perfectly matched layers (PMLs) are configured along the z-axis to effectively absorb outgoing radiation and simulate an open-space boundary. The complex refractive indices of the constituent materials are obtained from experimental data in the literature[32, 33]. To quantitatively characterize the AT performance, the isolation ratio (*IR*) of the transmittance is defined in decibels (dB) as:

$$IR = 10 \times \log_{10}\left(\frac{T_f}{T_b}\right) \quad (1)$$

where $T_f$ and $T_b$ represent the forward and backward transmittances, respectively. The *IR* is employed to characterize the transmissivity difference between the two propagation directions.

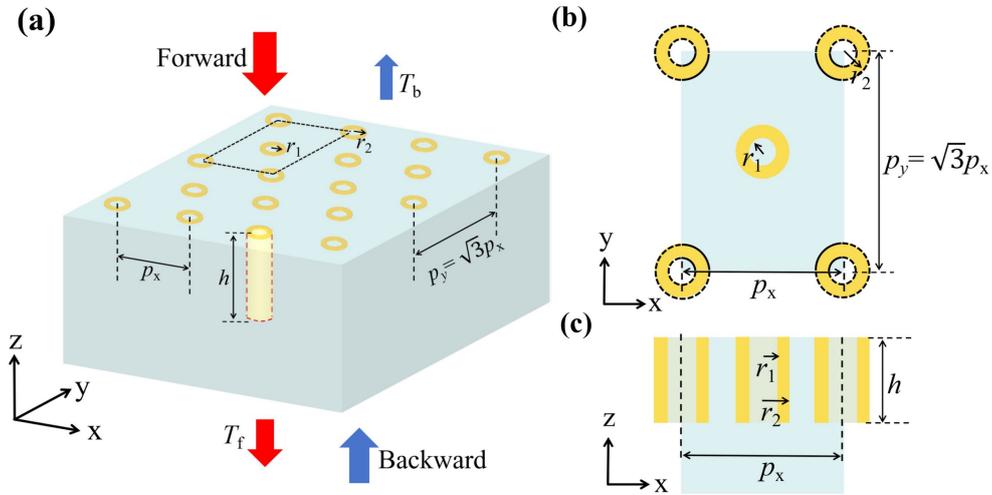

**Fig. 1.** Schematic of the periodic nanostructure. (a) 3D view. (b) Top view (xy plane) and (c) front view (xz plane) of the unit cell. The yellow and light-gray regions represent Si and SiO$_2$, respectively.

**Table 1. Parameters of the AT Devices**

| Parameters | unit: nm |
| --- | --- |
| Array period along x ($p_x$) | 2000 |
| Array period along y ($p_y$) | 3464.1 |
| Nanoring height ($h$) | 1100 |
| Inner radius ($r_1$) | 250 |
| Outer radius ($r_2$) | 440 |
| working band | 1750-2400 |

## 3. Results and discussion

As shown in Fig. 2**(a)**, for the optimized structure with an array period of $p_x$ = 2000 nm and $p_y$ = $\sqrt{3}p_x$, the device exhibits a remarkable broadband AT effect in the near-infrared region. The forward transmissivity remains consistently high, exceeding 0.8 across almost the entire simulated range (1750-2400 nm), with a peak value of approximately 0.98. In contrast, The backward transmissivity is significantly suppressed, falling below 0.15 across most of the wavelength range and reaching a minimum of near-zero around 2290 nm.

Figs. 2**(b)** shows the *IR* for forward and backward transmissivity at the array periods. Here, we define the AT band and its bandwidth of the device as the wavelength range in which *IR* is larger than 5 dB. The device demonstrates the wide bandwidth of approximately 650 nm ranging from roughly 1750 nm to 2400 nm. Specifically, the *IR* reaches its maximum value of 17.8 dB at 2280 nm. This broadband characteristic confirms that the proposed Si-based nanostructure can maintain high isolation efficiency while significantly expanding the operational wavelength range compared to conventional resonant structures.

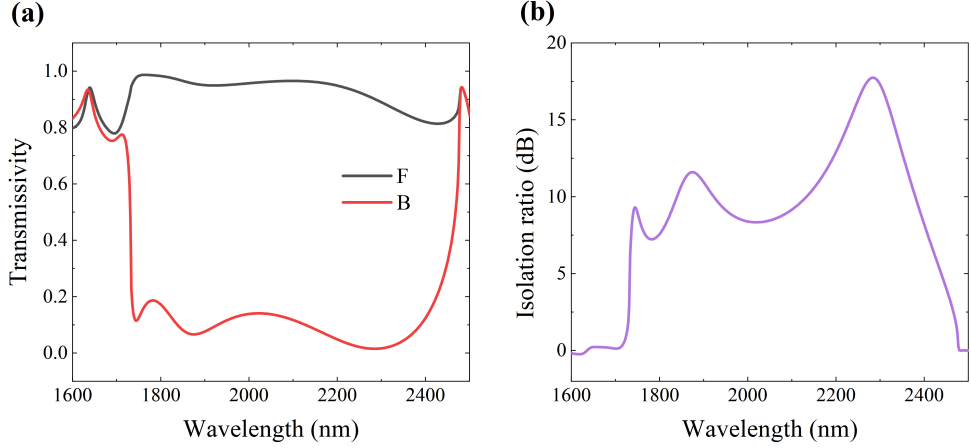

**Fig. 2.** Transmissivity performance of the nanostructure with $p_x$ = 2000 nm and $p_y$ = $\sqrt{3}\, p_x$. (a) Black and red curves stand for forward (F) and backward (B) transmissivities, respectively. (b) Calculated *IR* as a function of wavelength.

To clarify the mechanism of AT, Fig. 3 shows the diffraction-order contributions. For forward incidence, as shown in Fig. 3**(a)**, several diffraction channels are excited, and the total transmissivity results from the combined contributions of different diffraction orders. The zero-order component (0,0) remains very weak across the majority of the wavelength range from roughly 1750 nm to 2400 nm, indicating that it plays a negligible role in the forward transmissivity process. In contrast, higher-order diffraction channels dominate the transmissivity.

Specifically, the diffraction orders $T(\pm 1, \pm 1)$ contribute significantly across the entire broadband, maintaining a transmissivity level of approximately 0.2. Meanwhile, the $T(0, \pm 2)$ orders exhibit a marginal contribution, remaining consistently around 0.05 throughout most of the operation band. Notably, the combined effect of these higher-order channels leads to a high and stable total transmissivity exceeding 0.95 in the 1750-2400 nm range. For backward incidence, as depicted in Fig. 3**(b)**, all higher-order diffraction channels are effectively suppressed, and the total transmissivity is almost entirely determined by the zero-order component

(0,0). It is worth noting that the *T*(0,0) transmissivity under forward incidence almost overlaps with the total transmissivity under backward incidence. This confirms that the broadband AT originates from the efficient excitation of higher-order diffraction channels under forward illumination, while such channels are suppressed under backward illumination.

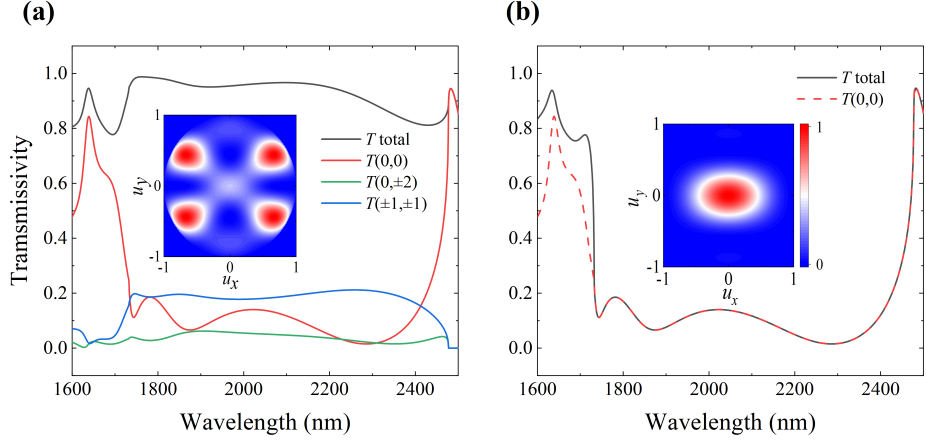

**Fig. 3.** Diffraction-induced transmissivity under (a) forward and (b) backward illuminations. *T* total and *T*(i, j) represent the total transmissivity and (i, j)-order diffraction-induced transmissivity, respectively. Insets represent the corresponding far-field transmissivity patterns of the structure at 2216 nm. $u_x$ and $u_y$ represent the normalized far-field projection coordinates along the x- and y-directions, respectively. The color bar indicates the normalized intensity, using the same scale for both inset figures.

This mechanism is further corroborated by the far-field radiation patterns shown in the insets of Figs. 3**(a)** and 3**(b)**. At the wavelength of 2216 nm, the forward far-field distribution exhibits four distinct off-axis intensity spots in the ($u_x$, $u_y$) plane, which correspond to the dominant ($\pm1,\pm1$) diffraction orders. Additionally, faint spots are also observed at the *T*(0, 0) and *T*(0, $\pm2$) positions. In contrast, the backward far-field pattern reveals only a single concentrated intensity spot at the center, representing the *T*(0, 0) mode. These far-field radiation characteristics are highly consistent with the spectral transmissivity behavior discussed above.

To demonstrate the energy transmissivity characteristics of the structure under forward and backward illumination, Fig. 4 shows the Poynting vector distributions.

The distributions are displayed on the cross-sectional planes at y = 0 nm and x = 0 nm. Under forward illumination, the electromagnetic energy propagates primarily along the incident direction, indicating that the energy effectively penetrates the structural layers. In contrast, under backward illumination, the energy is mainly reflected and confined to the structural units and the interstitial regions. The Poynting vector intensity above the substrate is relatively weak, suggesting that the energy transmissivity under backward illumination is negligible.

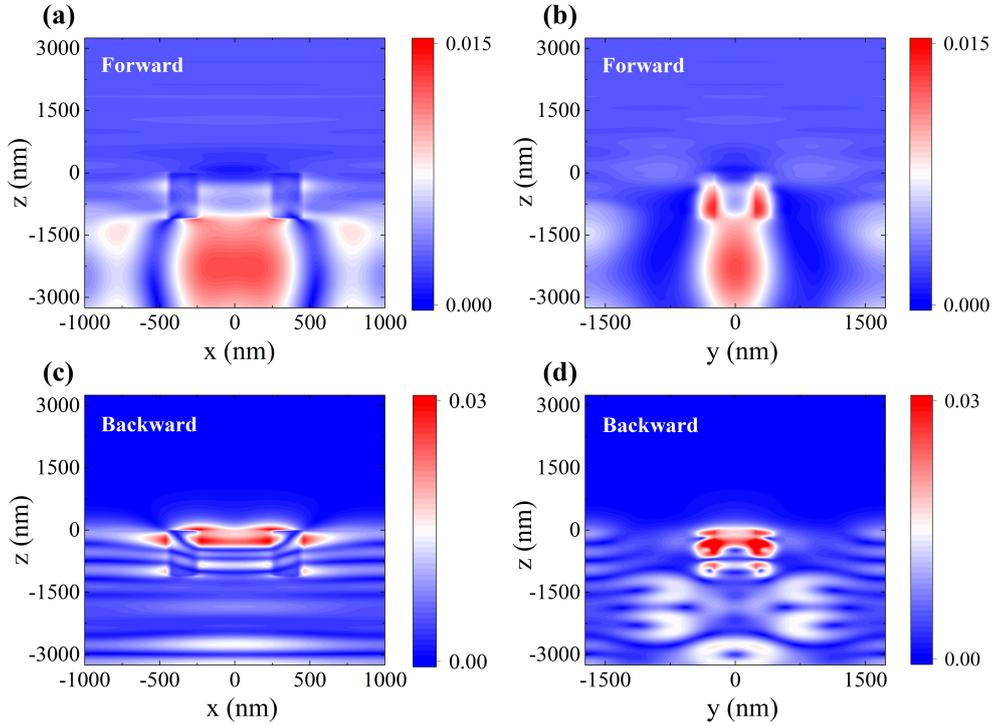

**Fig. 4.** The Poynting vector distributions for (a), (b) forward and (c), (d) backward illuminations at 2216 nm, in (a), (c) xz and (b), (d) yz planes. The color bar represents the intensity (arbitrary unit) of the Poynting vector.

To further verify the robustness and scalability of the proposed design, a uniform scaling factor $S$ ranging from 0.5 to 1.5 is applied to the original structure in Fig.5, including $r_1$, $r_2$, $h$, $p_x$, $p_y$. It is observed that the asymmetric transmissivity behavior is well preserved for all the above scaling cases, with consistently high forward transmissivity and suppressed backward transmissivity. Moreover, as the scaling factor $S$ increases, the transmissivity spectrum continuously redshifts, whereas a decrease in $S$ results in a corresponding blueshift. This spectral shift exhibits a continuous behavior, and the bandwidth gradually broadens as $S$ increases. Notably, such tunability can be achieved without altering the structural geometry, except for uniform scaling, providing a flexible approach for tailoring spectral responses over a broad wavelength range.

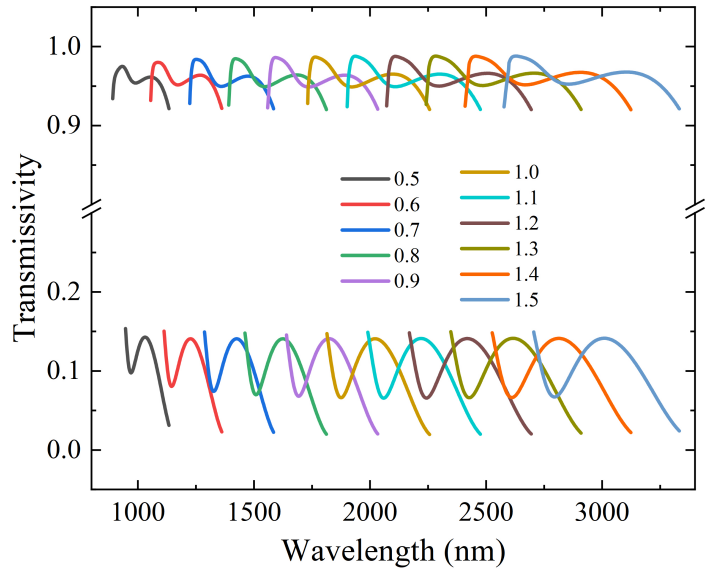

**Fig. 5.** Transmissivity spectra of the structure for different scaling factors $S$ ranging from 0.5 to 1.5.

Moreover, further calculations (not shown) indicate that the device exhibits polarization-insensitive behavior, i.e., the transmission spectra remain unchanged for different linear polarization angles of the incident light. This feature originates from the in-plane hexagonal symmetry of the structure. Owing to the geometric symmetry of the structure, which is defined by two basis vectors along the x-direction and 60° direction, and the absence of near-field coupling between neighboring nanorings (due to sufficiently large spacing), the optical response to arbitrarily polarized incident light can be linearly decomposed into the superposition of responses from these two polarization channels. This linear superposition principle renders the optical response polarization-insensitive, i.e., invariant under rotation of the incident linear polarization direction.

## 4. Conclusion

In summary, the AT device based on the embedded periodic Si nanoring array demonstrates excellent AT performance, with high forward and low backward transmissivities across wide ranges of wavelengths. By optimizing the geometric parameters of the nanorings, the device is specifically designed for the near-infrared region, exhibiting a robust broadband AT effect. For the optimized structure with a period of $p_x$ = 2000 nm and $p_y = \sqrt{3}p_x$, the AT band is located at 1750-2400 nm, with a significant bandwidth of approximately 650 nm. The maximum $IR$ reaches 17.8 dB at 2280 nm, while the forward transmissivity remains as high as 0.98. The mechanism underlying the broadband AT behavior originates from the excitation of multiple diffraction channels under forward illumination. Specifically, higher-order diffraction channels, primarily $T(\pm 1, \pm 1)$, contribute significantly to the forward transmissivity, whereas such channels are effectively suppressed under backward illumination. This mechanism is further corroborated by Poynting vector distributions, which reveal that electromagnetic energy penetrates the structural layers efficiently under forward illumination, while it is primarily reflected and confined to the structural units under backward illumination. Furthermore, the structure demonstrates excellent scalability through a uniform scaling factor $S$. As $S$ varies from 0.5 to 1.5, the transmissivity spectrum exhibits a continuous redshift, proving that the AT behavior is well preserved and can be tailored to various spectral ranges. These findings confirm that the Si-based nanoring structure is highly suitable for practical applications such as broadband optical filters, multi-channel optical sensors, and optical isolation.

## Acknowledgments

This work was supported by the National Natural Science Foundation of China (Grant No. 12504461).

## Disclosures

The authors declare no conflicts of interest.

## Data availability statement

The data that support the findings of this study are available upon reasonable request from the authors.

## References

1. H. T. Chen, A. J. Taylor and N. Yu, "A review of metasurfaces: physics and applications," Rep Prog Phys. **79**(7), 076401 (2016).
2. K. H. Kim, "Asymmetric Second-Harmonic Generation with High Efficiency from a Non-chiral Hybrid Bilayer Complementary Metasurface," Plasmonics. **16**(1), 77-82 (2021).
3. W. Liu, Z. Li, H. Cheng, *et al.*, "Dielectric Resonance-Based Optical Metasurfaces: From Fundamentals to Applications," iScience. **23**(12), 101868 (2020).
4. Z. L. Deng, Y. Cao, X. Li, *et al.*, "Multifunctional metasurface: from extraordinary optical transmission to extraordinary optical diffraction in a single structure," Photonics Res. **6**(5), 443 (2018).
5. X. Wei, Y. Sun, Y. Liang, *et al.*, "Multiband and bidirectional multiplexing asymmetric optical transmission empowered by nanograting-coupled defective multilayer photonic crystal," Sci Rep. **14**(1), 21190 (2024).
6. X. Ren, Y. Zhang, Y. Zhang, *et al.*, "Tunable Multi-Functional Metamaterial Based on Photosensitive Silicon for Unidirectional Reflectionlessness, Polarization Conversion, and Asymmetric Transmission," Materials. **18**(11), 2614 (2025).
7. A. Tripathi, C. F. Ugwu, V. S. Asadchy, *et al.*, "Nanoscale optical nonreciprocity with nonlinear metasurfaces," Nat Commun. **15**(1), 5077 (2024).
8. Y. Tian, Z. W. Chen, F. F. Ren, *et al.*, "High-Efficiency Asymmetric Transmission of Red-Near-Infrared Light Based on Chiral Metamaterial," Front. Phys. **9**, 676840 (2021).
9. S. Jahani and Z. Jacob, "All-dielectric metamaterials," Nat Nanotechnol. **11**(1), 23-36 (2016).
10. A. C. Overvig, S. Shrestha, S. C. Malek, *et al.*, "Dielectric metasurfaces for complete and independent control of the optical amplitude and phase," Light Sci Appl. **8**, 92 (2019).
11. A. S. Ansari, A. K. Iyer and B. Gholipour, "Asymmetric transmission in nanophotonics," Nanophotonics. **12**(14), 2639-2667 (2023).
12. J. H. Park, A. Ndao, W. Cai, *et al.*, "Symmetry-breaking-induced plasmonic exceptional points and nanoscale sensing," Nat. Phys. **16**(4), 462-468 (2020).
13. N. Parappurath, F. Alpeggiani, L. Kuipers, *et al.*, "The Origin and Limit of Asymmetric Transmission in Chiral Resonators," ACS Photonics. **4**(4), 884-890 (2017).
14. B. Y. Jin and C. Argyropoulos, "Self-Induced Passive Nonreciprocal Transmission by Nonlinear Bifacial Dielectric Metasurfaces," Phys. Rev. Appl. **13**(5), 054056 (2020).

15. Y. Cheng, K. Zhai, N. Zhu, *et al.*, "Optical non-reciprocity with multiple modes in the visible range based on a hybrid metallic nanowaveguide," J. Phys. D: Appl. Phys. **55**(19), 195102 (2022).
16. M. Montagnac, Y. Brule, A. Cuche, *et al.*, "Control of light emission of quantum emitters coupled to silicon nanoantenna using cylindrical vector beams," Light Sci Appl. **12**(1), 239 (2023).
17. A. Ozer, N. Yilmaz, F. T. Bagci, *et al.*, "Tunable and asymmetric transmission of light in visible spectrum," in *SPIE Photonics Europe* (SPIE2019), p. 110252D.
18. C. Argyropoulos, "Enhanced transmission modulation based on dielectric metasurfaces loaded with graphene," Opt Express. **23**(18), 23787-23797 (2015).
19. K. P. Dixit and D. A. Gregory, "Asymmetric nanocavity: from color-selective reflector to broadband near-infrared absorber," RSC Appl. Interfaces. **2**(4), 1059-1068 (2025).
20. M. Z. Chong, C. He, P. Feng, *et al.*, "Janus meta-imager: asymmetric image transmission and transformation enabled by diffractive neural networks," PhotoniX. **6**(1), 1-14 (2025).
21. J. Wang, T. Weber, A. Aigner, *et al.*, "Mirror-Coupled Plasmonic Bound States in the Continuum for Tunable Perfect Absorption," Laser Photonics Rev. **17**(11), 2300294 (2023).
22. C. Huang, Y. Feng, J. Zhao, *et al.*, "Asymmetric electromagnetic wave transmission of linear polarization via polarization conversion through chiral metamaterial structures," Phys. Rev. B. **85**(19), 8347-8356 (2012).
23. D. Conteduca, I. Barth, G. Pitruzzello, *et al.*, "Dielectric nanohole array metasurface for high-resolution near-field sensing and imaging," Nat Commun. **12**(1), 3293 (2021).
24. D. Hahnel, C. Golla, M. Albert, *et al.*, "A multi-mode super-fano mechanism for enhanced third harmonic generation in silicon metasurfaces," Light Sci Appl. **12**(1), 97 (2023).
25. M. Noman, H. Abutarboush, F. A. Tahir, *et al.*, "A novel multifunctional chiral metasurface with asymmetric transmission," Sci Rep. **14**(1), 24681 (2024).
26. N. Apurv Chaitanya, M. A. T. Butt, O. Reshef, *et al.*, "Lattice-plasmon-induced asymmetric transmission in two-dimensional chiral arrays," APL Photonics. **7**(1), 016105 (2022).
27. F. Yang, W. Cao, G. Zheng, *et al.*, "Plasmonic metasurfaces: Light-matter interactions, fabrication, applications and future outlooks," Prog. Mater. Sci. **154**, 101508 (2025).
28. W. W. Ahmed, H. Cao, C. Xu, *et al.*, "Machine learning assisted plasmonic metascreen for enhanced broadband absorption in ultra-thin silicon films," Light Sci Appl. **14**(1), 42 (2025).
29. H. Yang, C. Lou and X. Huang, "Broadband and highly efficient asymmetric optical transmission through periodic Si cylinder arrays on the dielectric substrates," Results Phys. **60**, 107691 (2024).
30. L. Zhang, J. Ding, H. Zheng, *et al.*, "Ultra-thin high-efficiency mid-infrared transmissive Huygens meta-optics," Nat Commun. **9**(1), 1481 (2018).

31. X. Hu, C. Xin, Z. Li, *et al.*, "Ultrahigh-contrast all-optical diodes based on tunable surface plasmon polaritons," New J. Phys. **12**(2), 023029 (2010).
32. D. F. Edwards, "Silicon (Si)," in Handbook of Optical Constants of Solids, E. D. Palik, ed. (Academic Press, 1997), pp. 547-569.
33. H. R. Philipp, "Silicon Dioxide ($SiO_2$) (Glass)," in Handbook of Optical Constants of Solids, E. D. Palik, ed. (Academic Press, 1997), pp. 749-763.